\documentclass[12pt]{article}

\def\beq{\begin{eqnarray}}
\def\eeq{\end{eqnarray}}

\def\L*{{\cal L}_*}
\def\lsim{\mathrel{\rlap{\lower3pt\hbox{\hskip0pt$\sim$}}
     \raise1pt\hbox{$<$}}}         
\def\gsim{\mathrel{\rlap{\lower4pt\hbox{\hskip1pt$\sim$}}
     \raise1pt\hbox{$>$}}}         

\usepackage{amsmath}
\usepackage{amsfonts}

\begin{document}
\begin{titlepage}

\centerline{\Large \bf Massive Gravity in Minkowski Space via}
\centerline{\Large \bf Gravitational Higgs Mechanism}
\medskip

\centerline{\large Zurab Kakushadze}

\bigskip

\centerline{\em 200 Rector Place, Apt 41F, New York, NY 10280}
\centerline{\tt zura@kakushadze.com}
\centerline{(October 15, 2007)}

\bigskip
\medskip

\begin{abstract}
{}In gravitational Higgs mechanism graviton components acquire mass in the process of spontaneous
breaking of diffeomorphisms (via scalar vacuum expectation values). Recently, gravitational Higgs mechanism was discussed
in the context of obtaining massive gravity directly in four dimensions. We discuss a setup, with all
diffeomorphisms spontaneously broken by scalars, where gravitational
Higgs mechanism results in massive gravity in Minkowski space with no non-unitary propagating modes. This is
achieved by including higher derivative terms in the scalar sector. Decoupling the non-unitary mode (trace of
the graviton) then requires tuning a single parameter, namely, the cosmological constant.
We also discuss a setup where only spatial diffeomorphisms are
broken. In this case we obtain massive gravity in Minkowski space with only transverse modes propagating,
so the mode count is the same as for a massless graviton.

\end{abstract}
\end{titlepage}

\newpage

\section{Introduction and Summary}

{}Gauge interactions are mediated by massless particles. Spontaneous breaking of gauge symmetry
results in gauge particles acquiring mass via Higgs mechanism. The massless particle associated with
general coordinate reparametrization invariance, the graviton, is then also expected to acquire mass
via gravitational Higgs mechanism \cite{KL} upon spontaneous breaking of diffeomorphism invariance.

{}In the context of domain wall backgrounds, where diffeomorphism invariance is broken partially,
gravitational Higgs Mechanism was originally discussed in \cite{KL}. One of the key points of \cite{KL}
is that in backgrounds with {\em spontaneously} (as opposed to explicitly) broken diffeomorphisms
extra modes can be gauged away using these diffeomorphisms as the
equations of motion are invariant under the full diffeomorphism invariance of the theory. For other related works,
{\em e.g.}, \cite{Duff, OP, GMZ, Perc, GT, Siegel, Por, AGS, Ch, Ban, Lec, Kir, Kiritsis, Tin, Ber}. For
a recent review of massive gravity in the context of infinite volume extra dimensions, see, {\em e.g.},
\cite{Gab} and references therein. For a recent review of spontaneous breaking of diffeomorphism symmetry
in the context of Lorentz violating Chern-Simons modification of gravity, see \cite{Jackiw} and references
therein.

{}Recently, 't Hooft revisited gravitational Higgs mechanism in the context of obtaining massive
gravity directly in four dimensions \cite{thooft}.
One of 't Hooft's motivations was not gravity itself, but QCD.
If QCD is to be described by string theory, all known consistent versions of which contain massless gravity,
then the graviton would somehow have to acquire mass (to describe massive glueball states).
Gravitational Higgs mechanism is one plausible way of achieving this.

{}In 't Hooft's setup four-dimensional diffeomorphisms are spontaneously broken by
four scalar fields whose VEVs are proportional to the four space-time coordinates. This is not
a static background as one of the four scalars is time-dependent. Einstein's equations have a flat
solution if a negative cosmological constant is introduced. Linearized gravity in this background is
massive, but one non-unitary mode (trace of the graviton) is also propagating. This is because
the massless graviton has two propagating degrees of freedom, while the massive one has five, so only
three space-like scalars can be ``eaten" in gravitational Higgs mechanism. So there is an extra non-unitary degree of
freedom, which does not decouple. This non-unitarity can be traced to one of the four scalars, the one
that breaks time-like diffeomorphisms, being (effectively) time-like. In \cite{thooft} a way of removing
this non-unitarity was discussed.

{}Based on the above count of propagating degrees of freedom, in \cite{ZK} a somewhat different approach was
pursued. Since, in four dimensions, the massless graviton has two propagating degrees of freedom,
and the massive graviton has five propagating degrees
of freedom, three space-like scalars should suffice for gravitational Higgs mechanism. In \cite{ZK} we discussed
a setup, where in $D$ dimensions $D-1$ space-like scalars spontaneously break {\em spatial} diffeomorphism invariance,
while the time-like diffeomorphisms are intact. The resulting background is no longer flat but linearly expanding.
Linearized gravity in this background is indeed massive with no non-unitary modes propagating, and the unbroken
time-like diffeomorphism is precisely what removes the non-unitary mode (trace of the graviton). An interesting feature
of the background discussed in \cite{ZK} is that the effective mass squared of the graviton is positive for $D<10$, it vanishes
in $D=10$, and it becomes tachyonic for $D>10$. It would be interesting to understand if there is a deeper reason
why classical considerations of \cite{ZK} single out $D=10$, the critical dimension of superstring theory.

{}In this note we discuss a setup where gravitational Higgs mechanism leads to massive gravity with no non-unitary propagating
modes in Minkowski space. This is achieved by simply including higher derivative terms in the scalar sector in the setup of
\cite{thooft}. The most general (Lorentz invariant) mass term for the graviton in the Lagrangian is of the form
\begin{equation}
 -{m^2\over 4} \left[h_{MN}h^{MN} - \beta (h^M_M)^2\right]~,
\end{equation}
where $\beta$ is a dimensionless parameter. Only for $\beta = 1$ does the trace component $h^M_M$ decouple. If we do not include
higher derivative terms in the scalar sector, then gravitational Higgs mechanism in the setup of \cite{thooft} invariably gives
$\beta = 1/2$. Once we include higher derivative terms, however, the resulting value of $\beta$ actually depends on higher derivative
couplings. We therefore need to tune one combination of couplings to get the desired value of $\beta = 1$. In fact, this turns out
to be nothing but tuning of the cosmological constant.

{}In this paper we consider a setup (which we discuss in detail in Sections 3, 4 and 5), where,
just as in the setup of \cite{thooft} (which we briefly review in Section 2),
all diffeomorphisms, including the time-like one, are spontaneously broken by
$D$ scalars, one of which is time-like. The scalar fluctuations are gauged away using the diffeomorphisms, while the trace component
decouples due to the aforementioned tuning of the cosmological constant. The mechanism for this decoupling is that at the tuned
value of the cosmological constant the kinetic term for scalar fluctuations reorganizes
into that of a vector boson with only $D-1$ propagating degrees of freedom
(all of which are unitary). This is why the resulting theory (after spontaneous breaking of diffeomorphisms) has the correct count of
$(D+1)(D-2)/2$ (and not $D(D-1)/2$) degrees of freedom, and is therefore unitary.

{}In section 6 we discuss an adaptation of the setup of \cite{ZK}, where only spatial diffeomorphisms are broken by $D-1$ space-like
scalars. Upon including higher derivative terms in the scalar sector, we find solutions with Minkowski background. It turns out that
in these solutions none of the scalar modes are propagating. As a result, we have only $D(D-3)/2$ propagating graviton modes, same as
in the massless case, except that these modes are actually massive\footnote{In such massive gravity models we do not expect
to have van Dam-Veltman-Zakharov \cite{vDV,Zak} discontinuity.}.

We end this note with a few brief remarks in Section 7.

\section{Spontaneous Breaking of Diffeomorphisms}

{}In this section we discuss the setup of \cite{thooft}. Consider the following action:
\begin{equation}
 S = M_P^{D-2}\int d^Dx \sqrt{-G}\left[ R - Z_{AB} \nabla^M \phi^A \nabla_M \phi^B - \Lambda\right]~,
 \label{actionphi1}
\end{equation}
where $\Lambda$ is the cosmological constant, and $Z_{AB}$ is a constant metric for the scalar sector, where
$A = 0,\dots,D-1$. In the following we will take this metric to be identical to the Minkowski metric: $Z_{AB} =
{\delta_A}^M{\delta_B}^N \eta_{MN}$.

{}The equations of motion read:
\begin{eqnarray}
 \label{phi11}
 && \nabla^2 \phi^A = 0~,\\
 \label{einstein11}
 && R_{MN} - {1\over 2}G_{MN} R = \nonumber\\
 &&Z_{AB} \left[\nabla_M\phi^A \nabla_N\phi^B
 -{1\over 2}G_{MN}Z_{AB} \nabla^S \phi^A \nabla_S \phi^B \right] -{1\over 2}G_{MN}\Lambda~.
\end{eqnarray}
There is a solution to this system with flat Minkowski metric \cite{OP, GMZ}:
\begin{eqnarray}\label{solphi}
 &&\phi^A = m~{\delta^A}_M~x^M~,\\
 \label{solG}
 &&G_{MN} = \eta_{MN}~,
\end{eqnarray}
where
\begin{equation}\label{mLambda}
 m^2 = -\Lambda / (D-2)~.
\end{equation}
So, the cosmological constant $\Lambda$ must be negative for such solutions to exist.

{}The scalar fluctuations $\varphi^A$ can be gauged away using the diffeomorphisms (under
$x^M \rightarrow x^M - \xi^M$):
\begin{equation}\label{diffphi1}
 \delta\varphi^A =\nabla_M \phi^A \xi^M = m~{\delta^A}_M ~\xi^M~.
\end{equation}
Note, however, that once we gauge away the scalars, diffeomorphisms
\begin{equation}
 \delta h_{MN} = \nabla_M \xi_N + \nabla_N \xi_M
\end{equation}
can no longer be used to gauge away any of the graviton components $h_{MN} \equiv G_{MN} - \eta_{MN}$.

{}After setting $\varphi^A = 0$, the linearized equations of motion (\ref{phi11}) and (\ref{einstein11}) read:
($G_{MN} = \eta_{MN} + h_{MN}$):
\begin{eqnarray}
 &&2\partial^N h_{MN} - \partial_M h = 0~,\\
 &&\partial_S\partial^S h_{MN} +\partial_M\partial_N
 h-\partial_M \partial^S h_{SN}-
 \partial_N \partial^S h_{SM}-\eta_{MN}
 \left[\partial_S\partial^S h-\partial^S\partial^R
 h_{SR}\right] = \nonumber\\
 \label{EOMh}
 &&m^2\left[2h_{MN} - \eta_{MN} h\right]~,
\end{eqnarray}
where $h\equiv h^M_M$.

{}We can now see the unitarity issue discussed in \cite{thooft}. To do this, let us consider a more
general set of equations:
\begin{eqnarray}\label{condh}
 &&\zeta\partial^N h_{MN} - \partial_M h = 0~,\\
 &&\partial_S\partial^S h_{MN} +\partial_M\partial_N
 h-\partial_M \partial^S h_{SN}-
 \partial_N \partial^S h_{SM}-\eta_{MN}
 \left[\partial_S\partial^S h-\partial^S\partial^R
 h_{SR}\right] = \nonumber\\
 &&m^2\left[\zeta h_{MN} - \eta_{MN} h\right]~,
 \label{EOMh-00}
\end{eqnarray}
where $\zeta$ is a parameter. Taking the trace of the second equation, we have:
\begin{equation}
 (D-2)\left[\partial^S\partial^R h_{SR} -  \partial^S\partial_S h\right] = -m^2 (D-\zeta)h~.
\end{equation}
On the other hand,
\begin{equation}
 \zeta \partial^M\partial^N h_{MN} = \partial^M\partial_M h~,
\end{equation}
so we have the following equation of motion for $h$:
\begin{equation}
 (D-2)(1-1/\zeta)\partial^S\partial_S h = m^2 (D-\zeta) h~.
\end{equation}
This means that, unless $\zeta = 1$, $h$ is a propagating degree of freedom, and since this degree of
freedom has negative norm, the corresponding theory is non-unitary. The number of degrees of freedom in
this model is $D(D+1)/2$ (from $h_{MN}$) less $D$ (from the condition (\ref{condh})), which gives $D(D-1)/2$.
This is massive gravity plus an {\em undecoupled} trace component $h$, a non-unitary theory. At the special
value of $\zeta = 1$ we have $h = 0$, and the number of propagating degrees of freedom is
$D(D-1)/2 - 1 = (D+1)(D-2)/2$, which is the number of degrees of freedom of a massive graviton. To get this count
of the propagating degrees of freedom we should have only $(D-1)$ space-like scalars ``eaten" in the gravitational
Higgs mechanism, not all $D$ scalars (which include the negative-norm time-like scalar). This approach was successfully
implemented in \cite{ZK}, where massive gravity with no propagating ghosts was obtained in a linearly expanding
background (see Sections 6 and 7).

{}Another way to view the reason for non-unitarity in this model is to note that we have an
indefinite metric $Z_{AB}$ for the scalar sector\footnote{Actually, in \cite{thooft} the metric $Z_{AB}$
is positive definite, but one of the scalars has imaginary VEV, $\phi^0 = imt$, with the same net effect.}.
Indeed, effectively, we just have a massless {\em vector}
meson $\phi^M$ with the Lagrangian $L\sim - \nabla^M \phi^N \nabla_M \phi_N$. This Lagrangian is clearly
non-unitary, hence the issue. In \cite{thooft} it was suggested that this non-unitarity can be removed by 
assuming a special form of the (tree-level) coupling for the non-unitary mode $h$ such that it does not couple 
to the matter energy-momentum tensor. This special form of the coupling is expected to be modified at the quantum level.

\section{Massive Gravity without Non-Unitary States}

{}To avoid the non-unitary mode, we need to effectively eliminate one of the scalars. In this section we argue that this
can be achieved by adding higher derivative terms in the scalar sector. Decoupling of the ghost then requires appropriately tuning the 
cosmological constant\footnote{This tuning can be done order-by-order in perturbative expansion, albeit since we are dealing with gravity, the theory is not
renormalizable, so this would have to be considered in a more fundamental framework such as string theory.}.

{}Thus, consider the induced metric for the scalar sector:
\begin{equation}
 Y_{MN} = Z_{AB} \nabla_M\phi^A \nabla_N\phi^B~.
\end{equation}
It is natural to generalize the action (\ref{actionphi1}) as follows:
\begin{equation}
 S_Y = M_P^{D-2}\int d^Dx \sqrt{-G}\left[ R - V(Y)\right]~,
 \label{actionphiY}
\end{equation}
where {\em a priori} $V(Y)$ is a generic function of $Y$, and
\begin{equation}\label{Y}
 Y\equiv Y_{MN}G^{MN}~.
\end{equation}
In particular, note that (\ref{actionphi1}) is a special case of (\ref{actionphiY}) with $V(Y) = \Lambda + Y$.

{}The equations of motion read:
\begin{eqnarray}
 \label{phiY}
 && \nabla^M\left(V^\prime(Y)\nabla_M \phi^A\right) = 0~,\\
 \label{einsteinY}
 && R_{MN} - {1\over 2}G_{MN} R = V^\prime(Y) Y_{MN}
 -{1\over 2}G_{MN} V(Y)~,
\end{eqnarray}
where prime denotes derivative w.r.t. Y.

{}We are interested in finding solutions of the form:
\begin{eqnarray}\label{solphiY}
 &&\phi^A = m~{\delta^A}_M~x^M~,\\
 \label{solGY}
 &&G_{MN} = \eta_{MN}~.
\end{eqnarray}
These exist for a class of ``potentials" $V(Y)$ such that the following equation
\begin{equation}\label{tune-V.0}
 Y~V^\prime(Y) = {D\over 2} ~V(Y)
\end{equation}
has non-trivial solutions. Let us denote such a solution by $Y_*$. Then we have
\begin{equation}
 m^2 = Y_* / D~.
\end{equation}
Note that the solution of the previous section given by (\ref{solphi}), (\ref{solG}) and (\ref{mLambda}) is
a special case of this more general class of solutions when the potential $V(Y)$ is linear.

{}As in the previous section, the scalar fluctuations $\varphi^A$ can be gauged away using the diffeomorphisms:
\begin{equation}\label{diffphiY}
 \delta\varphi^A =\nabla_M \phi^A \xi^M = m~{\delta^A}_M ~\xi^M~.
\end{equation}
However, once we gauge away the scalars, diffeomorphisms
can no longer be used to gauge away any of the graviton components $h_{MN}$. After setting $\varphi^A = 0$,
we have
\begin{eqnarray}
 && Y_{MN} = m^2\eta_{MN}~,\\
 && Y = m^2\eta_{MN}G^{MN} = m^2\left[D - h + \dots\right] = Y_* - m^2 h + \dots~,
\end{eqnarray}
where the ellipses stand for higher order terms in $h_{MN}$. The linearized equations of motion read:
\begin{eqnarray}
 &&2\partial^N h_{MN} V(Y_*) - \partial_M h \left[ V(Y_*) - 4m^4V^{\prime\prime}(Y_*)\right] = 0~,\\
 \label{linEOM}
 &&R_{MN} - {1\over 2}G_{MN} R = {1\over 4}\eta_{MN}h\left[V(Y_*) - 4m^4V^{\prime\prime}(Y_*)\right] -
 {1\over 2}h_{MN} V(Y_*)~.
\end{eqnarray}
Note that for a linear potential $V(Y)$ we reproduce the result of the previous section with the tensor structure
of the r.h.s. of (\ref{EOMh-00}). However, in the general case we have:
\begin{eqnarray}
 &&\partial^N h_{MN} - \beta \partial_M h  = 0~,\\
 \label{linEOM1}
 &&R_{MN} - {1\over 2}G_{MN} R = {M_h^2\over 2} \left(\beta\eta_{MN}h  - h_{MN} \right)~,
\end{eqnarray}
where
\begin{eqnarray}
 &&\beta \equiv {1\over 2} - 2m^4 {V^{\prime\prime}(Y_*)\over{V(Y_*)}}~,\\
 &&M_h^2\equiv V(Y_*)~.
\end{eqnarray}
In particular, for a special class of potentials with
\begin{equation}\label{tune-V}
 4Y_*^2 V^{\prime\prime}(Y_*) = -D^2 V(Y_*)~,
\end{equation}
we have $\beta = 1$, and the correct tensor structure for massive gravity without non-unitary states. In this case
the explicit form of the equations of motion is given by:
\begin{eqnarray}
 &&h = 0~,\\
 &&\partial^N h_{MN} = 0~,\\
 &&\partial^S\partial_S h_{MN} = M_h^2 h_{MN}~,
\end{eqnarray}
In particular, the ghost state $h$ decouples.

{}Thus, as we see, we can get the Pauli-Fierz combination of the mass terms
for the graviton if we tune {\em one} combination of couplings. In fact, this tuning is nothing but tuning of
the cosmological constant - indeed, (\ref{tune-V}) relates the cosmological constant to higher derivative couplings.

{}Thus, consider a simple example:
\begin{equation}
 V = \Lambda + Y + \lambda Y^2~.
\end{equation}
The first term is the cosmological constant, the second term is the kinetic term for the scalars (which can always be normalized
such that the corresponding coefficient is 1), and the third term is a four-derivative term. We then have:
\begin{equation}
 Y_* = -{D\over{2(D+2)}}~\lambda^{-1}~,
\end{equation}
which relates the mass parameter $m$ to the higher derivative coupling $\lambda$:
\begin{equation}
 m^2 = Y_* / D = -{1\over{2(D+2)}}~ \lambda^{-1}~.
\end{equation}
Note that we must have $\lambda < 0$. Moreover, the cosmological constant $\Lambda$ must be tuned as follows:
\begin{equation}
 \Lambda = {{D^2 + 4D - 8}\over {4(D+2)^2}}~\lambda^{-1}~,
\end{equation}
which implies that the cosmological constant must be negative.

\section{Why Does the Ghost Decouple?}

{}In the previous section we saw that at the special value of the cosmological constant the trace mode $h$ decouples, and
we have a unitary theory. This is in conflict with the following naive count of the degrees of freedom. We started with massless
gravity, whose number of degrees of freedom is given by $D(D-3)/2$, plus $D$ scalars, one of which is time-like. We therefore
expect $D(D-1)/2$ propagating degrees of freedom, $(D+1)(D-2)/2$ of which correspond to a massive graviton, and one to the ghost.
So, why does the ghost decouple?

{}The answer is {\em residual} gauge symmetry. As we will see in a moment, at the special value of the cosmological
constant the kinetic term for scalar fluctuations reorganizes
into that of a vector boson. To see this, let us not gauge away the scalar fluctuations. Let us
expand the action (\ref{actionphiY}) around the background (\ref{solphiY}) and (\ref{solGY}) keeping only linear and quadratic
terms in the scalar fluctuations $\varphi^a$ and the metric fluctuations $h_{MN}$. When (\ref{tune-V.0}) and (\ref{tune-V}) are
satisfied, we have the following expansion:
\begin{eqnarray}
 &&\sqrt{-G} V(Y) / V(Y_*) = {1\over 4}\left[h_{MN}h^{MN} - h^2\right] + {1\over m} \partial^M{\delta_M}^A\varphi_A +\nonumber\\
 &&{1\over m} \left[h \partial^M{\delta_M}^A\varphi_A - h^{MN} \partial_N \partial_S {\delta^S}_A \varphi^A\right] +\nonumber\\
 &&{1\over {2m^2}}\left[\partial^M\varphi^A\partial_M\varphi^B\eta_{AB} - \left(\partial^M{\delta_M}^A\varphi_A\right)^2\right] +\dots~,
\end{eqnarray}
where the ellipses stand for cubic and higher order terms. This then implies that, up to surface terms, we have the following action:
\begin{eqnarray}
 &&S_Y = M_P^{D-2}\int d^Dx \sqrt{-G}R - \nonumber\\
 &&M_P^{D-2}\int d^Dx \left[
 {M_h^2\over 4}\left(h_{MN}h^{MN} - h^2\right) + {M_h^2\over m} A^M Q_M +
 {M_h^2\over {4m^2}} F_{MN}F^{MN} \right]~,
\end{eqnarray}
where, as in the previous section, $M_h^2 \equiv V(Y_*)$, and
\begin{eqnarray}
 &&A_M \equiv {\delta_M}^A\varphi_A~,\\
 &&Q_M \equiv \partial^N h_{MN} - \partial_M h~,\\
 &&F_{MN} \equiv\partial_M A_N - \partial_N A_M~,
\end{eqnarray}
Thus, as we see, at the tuned value of the cosmological constant corresponding to (\ref{tune-V}), the kinetic term for scalar fluctuations reorganizes
into that of a vector boson, which is of the special form $F_{MN}F^{MN}$. Note that
the above action is invariant under the full diffeomorphisms:
\begin{eqnarray}
 &&\delta A_M = m\xi_M~,\\
 &&\delta h_{MN} = \partial_M\xi_N + \partial_N\xi_M~.
\end{eqnarray}
So, as in the previous section, we can gauge away all components of $A_M$. However, only $D-1$ of the $D$ components of $A_M$
are actually propagating in the above action. Indeed, the kinetic term $F_{MN}F^{MN}$ for the vector boson implies that only the
modes satisfying
\begin{equation}
 \partial^M A_M = 0
\end{equation}
have a pole in the propagator. In other words, in the appropriate coordinate frame,
the time-like component of $A_M$ is not propagating to begin with, and we can use
the spatial diffeomorphisms to set the space-like components of $A_M$ to zero. The remaining time-like diffeomorphism then can
be used to gauge away the ghost, {\em i.e.}, the trace component $h$. Covariantly, first consider $(D-1)$ diffeomorphisms satisfying
\begin{equation}
 \partial^M \xi_M = 0~.
\end{equation}
These diffeomorphisms can be used to gauge away $(D-1)$ components of $A_M$ such that $A_M$ can be written as
\begin{equation}
 A_M = \partial_M \chi~,
\end{equation}
where $\chi$ is a scalar. The action then reads:
\begin{eqnarray}
 &&S_Y = M_P^{D-2}\int d^Dx \sqrt{-G}R - \nonumber\\
 &&M_P^{D-2}\int d^Dx \left[
 {M_h^2\over 4}\left(h_{MN}h^{MN} - h^2\right) - {M_h^2\over m} \chi \partial^M Q_M \right]~,
\end{eqnarray}
so $\chi$ is not a propagating degree of freedom but a Lagrange multiplier leading to the constraint
\begin{equation}
 \partial^M Q_M = 0~.
\end{equation}
We still have one diffeomorphism remaining, which can be written as
\begin{equation}
 \xi_M = \partial_M \psi~.
\end{equation}
Under this diffeomorphism we have:
\begin{eqnarray}
 &&\delta\chi = m\psi~,\\
 &&\delta h_{MN} = 2\partial_M\partial_N\psi~,\\
 &&\delta h = 2\partial^S\partial_S \psi~.
\end{eqnarray}
This diffeomorphism only shifts the Lagrange multiplier $\chi$, and we can use it to gauge away the trace component $h$.

{}Let us see this in more detail. The equations of motion read:
\begin{eqnarray}
 &&\partial^S\partial^R h_{RS} - \partial^S\partial_S h = 0~,\\
 &&\partial_S\partial^S h_{MN} +\partial_M\partial_N
 h-\partial_M \partial^S h_{SN}-
 \partial_N \partial^S h_{SM}-\eta_{MN}
 \left[\partial_S\partial^S h-\partial^S\partial^R
 h_{SR}\right] = \nonumber\\
 &&M_h^2 \left[h_{MN} - \eta_{MN} h\right] - {2M_h^2\over m} \left[\partial_M \partial_N\chi - \eta_{MN} \partial^S\partial_S\chi\right]~.
 \label{EOMh-chi}
\end{eqnarray}
Differentiating the second equation w.r.t. $\partial^N$ and contracting indices, we obtain:
\begin{equation}
 \partial^N h_{MN} = \partial_M h~.
\end{equation}
Substituting this into (\ref{EOMh-chi}) we have:
\begin{equation}
 \partial^S\partial^R h_{RS} = M_h^2 h_{MN} + \partial_M\partial_N {\widetilde h}~,
\end{equation}
where
\begin{equation}
 {\widetilde h} \equiv h - {2M_h^2\over m} \chi~.
\end{equation}
Under the remaining diffeomorphisms $\psi$ we have the following transformation property:
\begin{equation}
 \delta {\widetilde h} = 2\left[\partial^S\partial_S\psi - M_h^2 \psi\right]~.
\end{equation}
This implies that we can always set ${\widetilde h}$ to zero:
\begin{equation}\label{gauge-h}
 {\widetilde h} = 0~.
\end{equation}
However, even after this gauge fixing, there is still {\em residual} gauge symmetry in the system. Thus, the gauge fixing
condition (\ref{gauge-h}) is preserved by diffeomorphisms satisfying the following massive Klein-Gordon equation:
\begin{equation}\label{psi-00}
 \partial^S\partial_S\psi - M_h^2 \psi = 0~,
\end{equation}
which is the same as that for the graviton modes:
\begin{equation}
 \partial^S\partial^R h_{RS} = M_h^2 h_{MN} ~.
\end{equation}
Recall that the trace component $h$ transforms as
\begin{equation}
 \delta h = 2\partial^S\partial_S \psi~.
\end{equation}
Thus, $h$ is a pure gauge and can be set to zero using the residual gauge transformations satisfying (\ref{psi-00}).
Then the propagating modes satisfy:
\begin{eqnarray}
 &&h = 0~,\\
 &&\partial^N h_{MN} = 0~,\\
 &&\partial^S\partial_S h_{MN} = M_h^2 h_{MN}~,
\end{eqnarray}
which describes massive gravity with $(D+1)(D-2)/2$ propagating degrees of freedom and no ghost.

{}Thus, as we see, the reason why the ghost decouples is that, at the tuned value of the cosmological constant, the number
of propagating scalar degrees of freedom is not $D$ but $D-1$. This is similar to the setup in \cite{ZK} where only
spatial diffeomorphisms are broken by space-like scalars. However, the difference between our setup here and in \cite{ZK}
is that here Lorentz invariance is explicitly preserved, and the reduction in the scalar degrees of freedom is due to
the fact that at the tuned value of the cosmological constant the kinetic term for scalar fluctuations reorganizes
into that of a vector boson.

\section{Massive Gravity via a Topological Coupling}

{}The results of the previous sections suggest that non-unitarity of the model of \cite{thooft} is simply
an artifact of restricting to the basic second-derivative action in the scalar sector. Once we add higher derivatives,
we can tune the cosmological constant such that we obtain a unitary massive theory. In hindsight this is not surprising.
Indeed, had we been able to obtain a unitary theory without tuning any coupling constants, it would imply that we could
obtain a consistent flat solution without any fine tuning directly in four dimensions. This would certainly be nice -- we
could then (almost) solve the cosmological constant problem -- however, it does not happen here.

{}In this section we discuss an alternative method of achieving the same result as in the previous section,
via a coupling of a topological nature. Thus, note that the following product
\begin{equation}\label{U}
 U\equiv {1\over {D!}}\epsilon^{M_1\dots M_D}\nabla_{M_1}\phi^{A_1}\dots\nabla_{M_D}\phi^{A_D}\epsilon_{A_1\dots A_D}
\end{equation}
is a scalar density of weight 1. Note that here $\epsilon^{M_1\dots M_D}$ is a tensor density of weight 1, while
$\epsilon_{A_1\dots A_D}$ is not a space-time tensor as it carries only the global $SO(d,1)$ indices. (\ref{U}) implies that
$U/\sqrt{-G}$ transforms as a scalar, just as does $Y$ from the previous section. In fact, note that
\begin{equation}
 U = \sqrt{-\det(Y_{MN})}~,
\end{equation}
and $Y_{MN}$ and $G_{MN}$ both have the same transformation properties.

{}We can therefore consider the following action (we will discuss a generalization thereof in the following):
\begin{equation}
 S_U = M_P^{D-2}\int d^Dx \left\{\sqrt{-G}\left[ R - Y - \Lambda\right] + \lambda U^2/\sqrt{-G} \right\}~,
 \label{actionphiYU}
\end{equation}
where we did not include a linear term in $U$ as it would be purely topological ({\em i.e.}, it would be independent of
the metric and not contribute into the graviton mass term). Also, $\Lambda$ is the cosmological constant, and $\lambda$ is a
higher derivative coupling constant.

{}Once again, we are interested in finding solutions of the form:
\begin{eqnarray}\label{solphiU}
 &&\phi^A = m~{\delta^A}_M~x^M~,\\
 \label{solGU}
 &&G_{MN} = \eta_{MN}~.
\end{eqnarray}
First, we will gauge away the scalar fluctuations just as in the previous section. Then,
let us expand the action (\ref{actionphiYU}) around this solution keeping only linear and quadratic terms in the graviton
fluctuations $h_{MN}$:
\begin{eqnarray}\label{linactionU}
 S_U = M_P^{D-2}\int d^Dx \left[{\cal L}_{\rm Lin} - a - bh - c_1h_{MN}h^{MN} - c_2h^2\right]~,
\end{eqnarray}
where hereafter ${\cal L}_{\rm Lin}$ is (up to total derivatives) the quadratic in $h_{MN}$ term in the expansion of $\sqrt{-G} R$ around flat Minkowski metric, and  
\begin{eqnarray}
 &&a = m^2(D - {\widetilde \Lambda} - {\widetilde \lambda})~,\\
 &&b = {m^2\over 2} (D-{\widetilde \Lambda} + {\widetilde \lambda} - 2)~,\\
 &&c_1 = {m^2\over 4}(4 - {\widetilde \lambda} + {\widetilde \Lambda} - D)~,\\
 &&c_2 = {m^2\over 8}(D - {\widetilde \Lambda} -{\widetilde \lambda} - 4)~,
\end{eqnarray}
and ${\widetilde \Lambda} \equiv -\Lambda / m^2$, ${\widetilde \lambda} \equiv \lambda m^{2(D-1)}$. The constant
term proportional to $a$ in the action (\ref{linactionU}) can be ignored. The linear term in $h$ must vanish (or else the
background is not a solution of the equations of motion):
\begin{equation}\label{tuneLambda}
 {\widetilde \Lambda} = D - 2 + {\widetilde \lambda}~.
\end{equation}
Our action then reads:
\begin{eqnarray}\label{linactionU1}
 S_U = M_P^{D-2}\int d^Dx \left[{\cal L}_{\rm Lin} - {m^2\over 2} h_{MN}h^{MN} + {m^2\over 4}(1 + {\widetilde\lambda}) h^2\right]~.
\end{eqnarray}
This gives the Pauli-Fierz action with graviton mass squared $M_h^2 = 2m^2$ if we take ${\widetilde \lambda} = 1$.

{}Thus, just as in the previous section, we see that we get the desired unitary theory if we tune the cosmological constant. Indeed,
the condition ${\widetilde \lambda} = 1$ fixes the parameter $m$ in terms of the higher derivative coupling $\lambda$:
\begin{equation}
 m^2 = \lambda^{-1/(D-1)}~.
\end{equation}
Then the condition (\ref{tuneLambda}) implies that
\begin{equation}
 \Lambda = -(D-1)m^2 = -(D-1)\lambda^{-1/(D-1)}~,
\end{equation}
{\em i.e.}, we must tune the cosmological constant (against a higher derivative coupling) to obtain a unitary theory. Just as in the
model of Section 3, in complete parallel with our discussion in Section 4, at the tuned value of the cosmological constant the kinetic term for scalar fluctuations reorganizes into that of a vector boson, and only $D-1$ components of which are propagating.

{}Note that the action (\ref{linactionU1}) leads to the equations of motion for $h_{MN}$ which, for generic values of ${\widetilde \lambda}$,
imply that
\begin{equation}\label{new-cond-hh}
 \partial^N h_{MN} = {1\over 2}(1 + {\widetilde\lambda}) \partial_M h~.
\end{equation}
Just as in Section 3, these conditions coincide with those arising as a result of the equations of motion for scalar fluctuations.
Thus, the scalar equations of motion read:
\begin{equation}\label{new-EOM-phi}
 \partial_M\left\{\left[\sqrt{-G}G^{MN} - {{\lambda} U^2\over{\sqrt{-G}}} E^{MN}\right]\partial_N\phi^A\right\} = 0~,
\end{equation}
where $E^{MN}$ is the inverse of $Y_{MN}$:
\begin{equation}
 Y_{MS}E^{SN} = {\delta_M}^N~,
\end{equation}
and we have used the identity
\begin{equation}
 {{\partial U}\over{\partial Y_{MN}}} =  {{\partial \sqrt{-\det(Y_{SR})}}\over{\partial Y_{MN}}} = {U\over 2} E^{MN}~.
\end{equation}
Expanding (\ref{new-EOM-phi}) to the linear order in $h_{MN}$ around the background (\ref{solphiU}) and (\ref{solGU})
and setting the scalar fluctuations to zero, we obtain (\ref{new-cond-hh}).

{}Let us note that, while the term proportional to $U^2$ in (\ref{actionphiYU}) might appear contrived, it can be recast
in a more familiar form by introducing a Lagrange multiplier field $\chi$:
\begin{equation}
 S_U = M_P^{D-2}\int d^Dx \left\{\sqrt{-G}\left[ R - Y - \Lambda - \chi^2 / \lambda\right] + 2\chi U \right\}~,
 \label{actionphiYU1}
\end{equation}
where the $\chi U$ coupling is topological. The $\chi$ field can be thought of as a scalar field in the infinite mass limit with a
topological coupling to the $\phi^A$ scalar sector.
Here we note that topological terms within the setup of \cite{thooft} were
discussed in a different context in \cite{Oda}.

{}Finally, let us note that the action (\ref{actionphiYU}) can be generalized as follows:
\begin{equation}
 S_{YU} = M_P^{D-2}\int d^Dx \sqrt{-G}\left[ R - V(Y,{\widetilde U})\right]~,
 \label{actionphiYU2}
\end{equation}
where ${\widetilde U}\equiv U/\sqrt{-G}$. The analysis in the general case completely parallels our discussion here, so we will not
repeat it.

\section{Spontaneous Breaking of Spatial\\
Diffeomorphisms}

{}In this section we discuss an adaptation of the setup of \cite{ZK}, where $D-1$ space-like scalars break spatial diffeomorphisms,
while the time-like diffeomorphism is intact. The background of \cite{ZK} is not static but expanding. In this section, by adding
higher derivative terms in the scalar sector we will obtain a Minkowski background.

{}Let $\phi^a$ be $d = D-1$ scalars, where the indices $a,b,\dots$ are raised and lowered by $\delta^{ab}$ and $\delta_{ab}$,
so the scalar sector possesses $SO(d)$ global symmetry. Let
\begin{eqnarray}
 &&Y_{MN} \equiv \nabla_M\phi^a \nabla_N\phi_a~,\\
 &&Y \equiv G^{MN}Y_{MN}~,\\
 &&U^M\equiv {1\over {d!}}\epsilon^{M N_1\dots N_d}\nabla_{N_1}\phi^{a_1}\dots\nabla_{N_d}\phi^{a_d}\epsilon_{a_1\dots a_d}~,\\
 &&U^2\equiv G_{MN}U^M U^N~.
\end{eqnarray}
Note that
\begin{eqnarray}
 \label{der-U}
 &&\nabla_M U^M = 0~,\\
 &&Y_{MN}U^N = 0~.
\end{eqnarray}
The second equation implies that the only local invariants we can construct from $Y_{MN}$ and $U^M$ (and not their derivatives)
are functions of $Y$ and $U^2$ only.

{}Consider the following action (we will consider a generalization thereof in the following):
\begin{equation}
 S_U = M_P^{D-2}\int d^Dx \left\{\sqrt{-G}\left[ R - Y - \Lambda\right] + \lambda U^2/\sqrt{-G} \right\}~.
 \label{actionphiYU3}
\end{equation}
We are interested in finding solutions of the form:
\begin{eqnarray}\label{solphiU.1}
 &&\phi^a = m~{\delta^a}_M~x^M~,\\
 \label{solGU.1}
 &&G_{MN} = \eta_{MN}~.
\end{eqnarray}
The Einstein equation reads (as before $-G$ is a shorthand for $-\det(G_{MN})$):
\begin{eqnarray}
 &&R_{MN} - {1\over 2}G_{MN} R = \nonumber\\
 &&Y_{MN} + \lambda G_{MS}G_{NR}U^S U^R(-G)^{-1} - {1\over 2}G_{MN}
 \left[Y + \Lambda + \lambda U^2 (-G)^{-1}\right]~.
\end{eqnarray}
Then to have a solution given by (\ref{solphiU.1}) and (\ref{solGU.1}), we must have:
\begin{eqnarray}\label{tune1Y}
 &&{\widetilde \Lambda} = D - 2~,\\
 \label{tune2Y}
 &&{\widetilde\lambda} = -1~,
\end{eqnarray}
where ${\widetilde \Lambda} \equiv -\Lambda / m^2$ and ${\widetilde \lambda} \equiv \lambda m^{2(d-1)}$. The first
condition (\ref{tune1Y}) relates $m^2$ to $\Lambda$. The second condition (\ref{tune2Y}) fixes the higher derivative coupling. However,
unlike in the model of the previous sections, this second condition does {\em not} come from requiring that the ghost decouple. Rather,
it is required to have a flat Minkowski background. The reason why two (as opposed to one, as in previous sections) conditions are
required to obtain a flat solution is that the scalar VEVs break Lorentz invariance, and two fields shift the background now, the trace
$h$ and the time-like component $\rho\equiv h_{00}$. Both must have vanishing linear terms in the corresponding expansion of the
action, hence two conditions\footnote{Here we note that if we do not insist on the Minkowski background, one condition can
be relaxed. In this case, however, we have an expanding background as in \cite{ZK}, where ${\widetilde \Lambda} = {\widetilde \lambda} = 0$.}.

{}Let us now expand our action (\ref{actionphiYU3}) keeping only linear and quadratic terms in the scalar fluctuations $\varphi^i\equiv
{\delta^i}_a\varphi^a$ and metric fluctuations $h_{MN}$. Actually, the linear terms in the metric fluctuations vanish due to
(\ref{tune1Y}) and (\ref{tune2Y}), while the linear
terms in scalar fluctuations only come in total derivatives. Up to surface terms we have the following action:
\begin{eqnarray}
 &&S_U = M_P^{D-2}\int d^Dx ~{\cal L}_{\rm Lin} - \nonumber\\
 &&M_P^{D-2}\int d^Dx \left[
 {m^2\over 2}\left(h_{ij}h^{ij} - {\overline h}^2\right) + 2m \varphi^i Q_i +
 {1\over 2} F_{ij}F^{ij} \right]~,
 \label{linactionU1.2}
\end{eqnarray}
where
\begin{eqnarray}
 &&{\overline h} \equiv h^i_i~,\\
 &&Q_i\equiv \partial^j h_{ij} - \partial_i {\overline h}~,\\
 &&F_{ij} \equiv\partial_i \varphi_i - \partial_j \varphi_i~.
\end{eqnarray}
Thus, we have a Pauli-Fierz type of mass term, but only for spatial components of the graviton, and the Lorentz invariance $SO(d,1)$,
not surprisingly, is broken down to $SO(d)$, which is consistent with the gauge choice (\ref{solphiU.1}).

{}The action (\ref{linactionU1.2}) is invariant under the full diffeomorphisms:
\begin{eqnarray}
 &&\delta\varphi_i = m\xi_i~,\\
 &&\delta h_{MN} = \partial_M\xi_N + \partial_N\xi_M~.
\end{eqnarray}
So, we can gauge away the scalars $\varphi^i$. However, note that the action (\ref{linactionU1.2}) does not contain any time
derivatives of $\varphi^i$. This implies that the scalar fluctuations are actually {\em not} propagating in this background. We
therefore expect only $D(D-3)/2$ propagating graviton modes, same as for a massless graviton.

{}To proceed further, let us simplify our action by gauge fixing the scalars. However, instead of setting them all to zero, we
will set only $d-1 = D-2$ of them to zero using the diffeomorphisms satisfying the condition
\begin{equation}
 \partial^i \xi_i = 0~.
\end{equation}
Then we can bring $\varphi^i$ to the following form,
\begin{equation}
 \varphi_i = \partial_i {\overline \varphi}~,
\end{equation}
where ${\overline \varphi}$ is a single scalar field. The action now reads:
\begin{eqnarray}
 &&S_U = M_P^{D-2}\int d^Dx ~{\cal L}_{\rm Lin} - \nonumber\\
 &&M_P^{D-2}\int d^Dx \left[
 {m^2\over 2}\left(h_{ij}h^{ij} - {\overline h}^2\right) - 2m {\overline \varphi} \partial^i Q_i\right]~.
 \label{linactionU1.3}
\end{eqnarray}
So ${\overline \varphi}$ is just a Lagrange multiplier. Note that so far we used up only $D-2$ diffeomorphisms. Under
the remaining two diffeomorphisms $\xi_0$ and $\psi$, where $\xi_i \equiv \partial_i\psi$, we have:
\begin{eqnarray}
 &&\delta {\overline \varphi} = m \psi~,\\
 &&\delta \rho = 2 \xi_0^\prime~,\\
 &&\delta A_i = \partial_i \left[\xi_0 + \psi^\prime\right]~,\\
 &&\delta h_{ij} = 2\partial_i\partial_j \psi~,\\
 &&\delta{\overline h} = 2\partial^i\partial_i\psi~,
\end{eqnarray}
where $\rho \equiv h_{00}$, $A_i \equiv h_{0i}$, and prime denotes derivative w.r.t. time $t = x^0$. Using these
diffeomorphisms we can therefore gauge away $\rho$ and ${\overline h}$. However, we will keep them both for now.

{}Let us analyze the equations of motion in detail:
\begin{eqnarray}
 &&\partial^i\partial^j h_{ij} - \partial^i\partial_i {\overline h} = 0~,\\
 &&\partial_S\partial^S h_{MN} +\partial_M\partial_N h-\partial_M \partial^S h_{SN}-
 \partial_N \partial^S h_{SM}-\eta_{MN}
 \left[\partial_S\partial^S h-\partial^S\partial^R
 h_{SR}\right] = \nonumber\\
 \label{EOMh-spatial}
 &&{\delta_M}^i {\delta_N}^j \left[2m^2\left(h_{ij} - \delta_{ij} {\overline h}\right) -
 4m\left(\partial_i\partial_j{\overline \varphi} - \delta_{ij}\partial^k\partial_k {\overline\varphi}\right)\right]~,
\end{eqnarray}
where the first equation is the ${\overline \varphi}$ equation of motion. We have:
\begin{equation}
 \partial^S\partial^R h_{SR} - \partial^S\partial_S h = {\overline h}^{\prime\prime} + \partial^i\partial_i \rho - 2\partial^i A_i^\prime~.
\end{equation}
Moreover, differentiating (\ref{EOMh-spatial}) w.r.t. $\partial^N$ and contracting indices, we obtain:
\begin{equation}\label{hh}
 \partial^j h_{ij} = \partial_i {\overline h}~.
\end{equation}
The $(00)$ component of (\ref{EOMh-spatial}) is automatically satisfied. The $(0i)$ and the $(ij)$ components read (here we take
into account (\ref{hh})):
\begin{eqnarray}
 \label{EOM-0i}
 &&\partial^j\partial_j A_i - \partial_i\partial^j A_j = 0~,\\
 &&\partial^k\partial_k h_{ij} - h_{ij}^{\prime\prime} - \partial_i\partial_j\rho - \partial_i\partial_j {\overline h} +
 \partial_i A^\prime_j + \partial_j A^\prime_i + \delta_{ij}\left[{\overline h}^{\prime\prime} + \partial^k\partial_k\rho -
 2\partial^k A_k^\prime\right] = \nonumber\\
 &&2m^2\left(h_{ij} - \delta_{ij} {\overline h}\right) -
 4m\left(\partial_i\partial_j{\overline \varphi} - \delta_{ij}\partial^k\partial_k {\overline\varphi}\right)~.
\end{eqnarray}
Note that we can always write $A_i$ as follows:
\begin{eqnarray}
 && A_i = {\widetilde A}_i + \partial_i \omega~,\\
 && \partial^i {\widetilde A}_ i = 0~.
\end{eqnarray}
Then, according to (\ref{EOM-0i}), we have:
\begin{equation}
 \partial^j\partial_j {\widetilde A}_i = 0~,
\end{equation}
which implies that ${\widetilde A}_i = 0$, and we have
\begin{equation}
 A_i = \partial_ i\omega~.
\end{equation}
With this simplification, our equation of motion for $h_{ij}$ now reads:
\begin{eqnarray}
 &&\partial^k\partial_k h_{ij} - h_{ij}^{\prime\prime} - \delta_{ij}\left[\partial^k\partial_k {\overline h} - {\overline h}^{\prime\prime} \right] = \nonumber\\
 &&2m^2\left(h_{ij} - \delta_{ij} {\overline h}\right) +
 \left(\partial_i\partial_j{\overline \rho} - \delta_{ij}\partial^k\partial_k {\overline\rho}\right)~,
\end{eqnarray}
where
\begin{equation}
 {\overline \rho} \equiv \rho + {\overline h} - 2\omega^\prime - 4m {\overline \varphi}~.
\end{equation}
Note that under the remaining diffeomorphisms $\xi_0$ and $\psi$, we have
\begin{equation}
 \delta {\overline \rho} = 2\left[\partial^i\partial_i\psi - \psi^{\prime\prime} - 2m^2\psi\right]~,
\end{equation}
so ${\overline \rho}$ actually is invariant under the time-like diffeomorphism $\xi_0$.
However, we can gauge ${\overline \rho}$ away using the $\psi$ diffeomorphism:
\begin{equation}\label{rho.0}
 {\overline \rho} = 0~.
\end{equation}
We then have the following equation of motion for $h_{ij}$
\begin{equation}
 \partial^k\partial_k h_{ij} - h_{ij}^{\prime\prime} - 2m^2 h_{ij} = 0~.
\end{equation}
Note that the trace component ${\overline h}$ is non-unitary. However, as we will see in a moment, it actually does not propagate.

{}The reason for this is {\em residual} gauge symmetry. Indeed, the gauge condition (\ref{rho.0}) is preserved by diffeomorphisms
$\psi$ satisfying the following massive Klein-Gordon equation:
\begin{equation}
 \partial^i\partial_i\psi - \psi^{\prime\prime} - 2m^2\psi = 0~,
\end{equation}
which is the same as that for the graviton modes. On the other hand, we have the following transformation property:
\begin{equation}\label{psi.0}
 \delta {\overline h} = 2\partial^k\partial_k \psi~.
\end{equation}
Thus, $h$ is a pure gauge, and can be set to zero using the residual $\psi$ gauge transformations satisfying (\ref{psi.0}):
${\overline h} = 0$. This is
similar to how the ghost component decouples in the non-static background of \cite{ZK}, as well as in the Minkowski background in
Section 4.

{}Thus, the only propagating components are the traceless part of $h_{ij}$:
\begin{eqnarray}
 &&h^i_i = 0~,\\
\label{tansverse-cond.0}
 &&\partial^j h_{ij} = 0~,\\
 &&\partial^k\partial_k h_{ij} - h_{ij}^{\prime\prime} = 2m^2 h_{ij}~.
\end{eqnarray}
So we have $D(D-3)/2$ massive components, the same count as for a massless graviton, which is due to the additional $D-1$ transversality conditions
(\ref{tansverse-cond.0}).

{}Finally, let us mention a generalization of this setup.
Let
\begin{equation}
 W \equiv G_{MN} U^M U^N / (-\det(G_{MN}))~.
\end{equation}
Then a general action constructed from $Y$ and $W$ (but not their derivatives) reads:
\begin{equation}
 S_{YW} = M_P^{D-2}\int d^Dx \sqrt{-G}\left[ R - V(Y, W) \right]~,
 \label{actionphiYW}
\end{equation}
where $V(Y, W)$ is a generic function. In this section we considered a special case where $V(Y, W) = \Lambda + Y - \lambda W$.
The analysis in the general case completely parallels our
discussion here, so we will not repeat it. Let us mention, however, that the Lorentz-violating graviton mass term in the Lagrangian in these cases
is more general and takes the form
\begin{equation}
 -{M^2\over 4}\left(h_{ij}h^{ij} - \beta {\overline h}^2\right)~,
\end{equation}
where $\beta$ depends on higher dimensional couplings involving $Y$ and $W$. Similar models were discussed in \cite{Grisa}, except that in our case
diffeomorphisms are broken spontaneously (as opposed to explicitly).

\section{Comments}

{}The approach we took in this note to decoupling the trace component $h$ is somewhat different from that in \cite{thooft}. As we saw in the previous sections,
in the model of Section 2 the non-unitary mode $h$ did not decouple due to the absence of higher derivative terms. Once we add the
latter, we can decouple $h$ at the cost of tuning the cosmological constant.

{}An alternative approach was proposed in \cite{ZK}: It is to have only spatial diffeomorphisms broken. In this case the theory is
explicitly unitary. As we saw in Section 6, we can construct solutions with the Minkowski background in this setup if we add higher derivative terms.
An interesting feature of these solutions is that the scalar modes are not propagating, and the resulting massive graviton has only
transverse propagating modes, the same as in the massless case.

{}In \cite{GG} it was argued that simply adding a mass term for the graviton,
even in the Pauli-Fierz combination, makes the theory unstable, and that
stable theories of massive gravity must be either non-local or obtained via higher dimensional embeddings. The theories we discussed in this
note are local field theories. If we integrate the scalars out, then, in terms of pure gravity, the resulting theory is indeed non-local. It is
possible to embed massive gravity within local field theories (with additional fields) because
diffeomorphisms are not broken explicitly ({\em e.g.}, by simply adding a Pauli-Fierz term), but only spontaneously.
Cosmological stability of such models may not even be an issue for 't Hooft's QCD motivation (see Introduction).

{}Finally, let us remark that, in the context of open string field theory, Higgs mechanism for spin-2 states was mentioned in \cite{GT},
and discussed in \cite{Siegel}\footnote{I would like to thank Warren Siegel for pointing this out to me.}. In analogy with the massive spin-1
case, to obtain the scalar action in the spin-2 case, one can start from the Pauli-Fierz term and work backwards by applying diffeomorphisms to the mass
term. The gauge parameters then are interpreted as new scalar fields. This procedure implies that not only two- but also four-derivative terms are
required, however, it does not uniquely fix higher derivative terms\footnote{In \cite{Siegel} it was suggested that, to obtain the Pauli-Fierz term,
the four-derivative term in the scalar action is required be the Skyrme term \cite{Skyrme} for the translation group. One property of the Skyrme term is that
it contains only two time derivatives, which is important for unitarity in the global case.
However, as we saw in Section 3, the Pauli-Fierz term can be obtained for other (non-Skyrme) four-derivative actions as well, and
what is relevant for unitarity is not the number of time derivatives, but the non-propagation of the time-like scalar fluctuations. This is because here we are
dealing with a locally symmetric (diffeomorphism invariant) theory.}. In fact, as we saw in this paper, there is only one condition we need to impose to obtain
the Pauli-Fierz term, namely, that on the cosmological constant.

{}In this paper we took a different route. We started with diffeomorphism invariant actions with higher derivatives for scalars, and discussed what
happens upon spontaneous breaking of diffeomorphisms. This led us to the conclusion that, to obtain a unitary (Pauli-Fierz) mass term, the cosmological
constant needs to be tuned such that the time-like scalar fluctuations do not propagate (and the ghost decouples).



\begin{thebibliography}{99}

\bibitem{KL} Z. Kakushadze and P. Langfelder,
``Gravitational Higgs Mechanism", Mod. Phys. Lett. A15 (2000) 2265, arXiv:hep-th/0011245.

\bibitem{Duff} M.J. Duff,
``Dynamical Breaking of General Covariance and Massive Spin-2 Mesons", Phys. Rev. D12 (1975) 3969.

\bibitem{OP} C. Omero and R. Percacci,
``Generalized Nonlinear Sigma Models In Curved Space And Spontaneous Compactification", Nucl. Phys. B165 (1980) 351.

\bibitem{GMZ} M. Gell-Mann and B. Zwiebach,
``Space-Time Compactification Due To Scalars", Phys. Lett. B141 (1984) 333.

\bibitem{Perc} R. Percacci, 
``The Higgs Phenomenon in Quantum Gravity", Nucl. Phys. B353 (1991) 271.

\bibitem{GT} M.B. Green and C.B. Thorn,
``Continuing between Closed and Open Strings", Nucl. Phys. B367 (1991) 462.

\bibitem{Siegel} W. Siegel,
``Hidden Gravity in Open-String Field Theory", Phys. Rev. D49 (1994) 4144, arXiv:hep-th/9312117.

\bibitem{Por} M. Porrati,
``Higgs Phenomenon for 4-D Gravity in Anti de Sitter Space", JHEP 0204 (2002) 058, arXiv:hep-th/0112166.

\bibitem{AGS} N. Arkani-Hamed, H. Georgi and M.D. Schwartz,
``Effective Field Theory for Massive Gravitons and Gravity in Theory Space", Annals Phys. 305 (2003) 96, arXiv:hep-th/0210184.

\bibitem{Ch} A.H. Chamseddine,
``Spontaneous Symmetry Breaking for Massive Spin-2 Interacting with Gravity", Phys. Lett. B557 (2003) 247, arXiv:hep-th/0301014.

\bibitem{Ban} I.A. Bandos, J.A. de Azcarraga, J.M. Izquierdo and J. Lukierski,
``Gravity, p-branes and a Space-time Counterpart of the Higgs Effect", Phys. Rev. D68 (2003) 046004, arXiv:hep-th/0301255.

\bibitem{Lec} M. Leclerc,
``The Higgs Sector of Gravitational Gauge Theories", Annals Phys. 321 (2006) 708, arXiv:gr-qc/0502005.

\bibitem{Kir} I. Kirsch,
``A Higgs Mechanism for Gravity", Phys. Rev. D72 (2005) 024001, arXiv:hep-th/0503024.

\bibitem{Kiritsis} E. Kiritsis,
``Product CFTs, Gravitational Cloning, Massive Gravitons and the Space of Gravitational Duals", JHEP 0611 (2006) 049, arXiv:hep-th/0608088.

\bibitem{Tin} M.V. Bebronne and P.G. Tinyakov,
``Massive Gravity and Structure Formation", Phys. Rev. D76 (2007) 084011, arXiv:0705.1301 [astro-ph].

\bibitem{Ber} Z. Berezhiani, D. Comelli, F. Nesti and L. Pilo, 
``Spontaneous Lorentz Breaking and Massive Gravity", Phys. Rev. Lett. 131101 (2007) 99, arXiv:hep-th/0703264.

\bibitem{Gab} G. Gabadadze, 
``Cargese Lectures on Brane Induced Gravity", arXiv:0705.1929 [hep-th].

\bibitem{Jackiw} R. Jackiw, 
``Lorentz Violation in Diffeomorphism Invariant Theory", arXiv:0709.2348 [hep-th].

\bibitem{thooft} G. 't Hooft, 
``Unitarity in the Brout-Englert-Higgs Mechanism for Gravity", arXiv:0708.3184 [hep-th].

\bibitem{ZK} Z. Kakushadze, 
``Gravitational Higgs Mechanism and Massive Gravity", arXiv:0709.1673 [hep-th].

\bibitem{vDV} H. van Dam and M.J.G. Veltman, 
``Massive and Massless Yang-Mills and Gravitational Fields", Nucl. Phys. B22 (1970) 397.

\bibitem{Zak} V. I. Zakharov, 
``Linearized Gravitation Theory and the Graviton Mass", JETP Lett. 12 (1970) 312.

\bibitem{Oda} I. Oda,
``Gravitational Higgs Mechanism with a Topological Term", arXiv:0709.2419 [hep-th].

\bibitem{Grisa} G. Gabadadze and L. Grisa,
"Lorentz-violating Massive Gauge and Gravitational Fields", Phys. Lett. B609 (2005) 167,
arXiv:hep-th/0411278.

\bibitem{GG} G. Gabadadze and A. Gruzinov,
``Graviton Mass or Cosmological Constant?", Phys. Rev. D72 (2005) 124007, arXiv:hep-th/0312074.

\bibitem{Skyrme} T.H.R. Skyrme,
``A Nonlinear Field Theory", Proc. Roy. Soc. Lond. A260 (1961) 127.


\end{thebibliography}
\end{document}